\documentclass[aps,pra,reprint,superscriptaddress,longbibliography,floatfix]{revtex4-2}
\usepackage[T1]{fontenc}
\usepackage[utf8]{inputenc}
\usepackage{amsmath,amssymb,bm,graphicx,mathtools}
\usepackage{physics}
\usepackage{booktabs}
\usepackage{xcolor}
\usepackage{hyperref}
\hypersetup{colorlinks=true,linkcolor=blue,citecolor=blue,urlcolor=blue}

\begin{document}

\title{Normal forms of unidirectional coupling in quasi-phase-matched non-Hermitian systems}

\author{Kestutis Staliunas}
\affiliation{ICREA, Passeig Llu\'is Companys 23, 08010 Barcelona, Spain}
\affiliation{UPC, Departament de F\'isica, Rambla Sant Nebridi 22, 08222 Terrassa (Barcelona), Spain}

\begin{abstract}
Optimal conditions for unidirectional coupling in quasi-phase-matched non-Hermitian systems are analyzed for both autonomous and externally driven configurations. The quasi-phase-matched coupling mechanism is due to periodic modulation of the real and imaginary parts of the coupling interface, which results in unequal coupling coefficients between interacting waves. The conventional parity-time ($\mathcal{PT}$) symmetry theory suggests that the strongest unidirectionality should occur exactly at the exceptional point (EP). We show that this expectation is generally incorrect, as the optimum is shifted away from the EP depending on the detuning from exact quasi-phase-matching resonance. We formulate a unified two-mode description for autonomous and driven systems, and derive the corresponding normal forms near the shifted singularities associated with the EPs. 
\end{abstract}

\maketitle

\section{Introduction}

Unidirectional response in time is one of the clearest physical signatures of causality. Causality requires the response function to vanish for future times, and the analyticity of the corresponding susceptibility in the upper complex-frequency plane yields the Kramers-Kronig relations between its real and imaginary parts \cite{Kronig1926,Kramers1927,Toll1956,Nussenzveig1972}. A closely analogous concept can be realized in space as well. If the Fourier spectrum of a susceptibility vanishes on a half-axis of spatial wave numbers, the real and imaginary parts of the susceptibility are again related by a Hilbert transform (the spatial analogue of the Kramers--Kronig relation), now in the spatial coordinate \cite{Horsley2015,Longhi2015,Longhi2016}. This observation led to a broad family of one-sided scattering structures including reflectionless or invisible media, generalized non-Hermitian gratings, and unidirectional coupling devices \cite{Lin2011,Regensburger2012,Feng2013,Mostafazadeh2013,Zhao2018,Zhang2021,Zheng2023}. The spatial implementation is particularly attractive for photonics because the required non-Hermitian profiles can be realized through a controlled combination of refractive-index and gain/loss modulations.

The relation between Parity-Time ($\mathcal{PT}$) symmetry, Hilbert-transform media, and directionality fields has been developed in several complementary ways. In locally $\mathcal{PT}$-symmetric but globally parity-symmetric systems, the balance between gain/loss and refractive modulation can be imposed only in selected spatial regions, showing that the local complex structure rather than a global symmetry operation controls the directional response \cite{Ahmed2016}. An even more direct connection is provided by the local Hilbert-transform approach of Ref.~\cite{Ahmed2018}, where prescribed directionality fields are generated by constructing non-Hermitian potentials whose real and imaginary parts satisfy a local Hilbert-transform condition. 

\begin{figure}[t]
    \includegraphics[width=1\columnwidth]{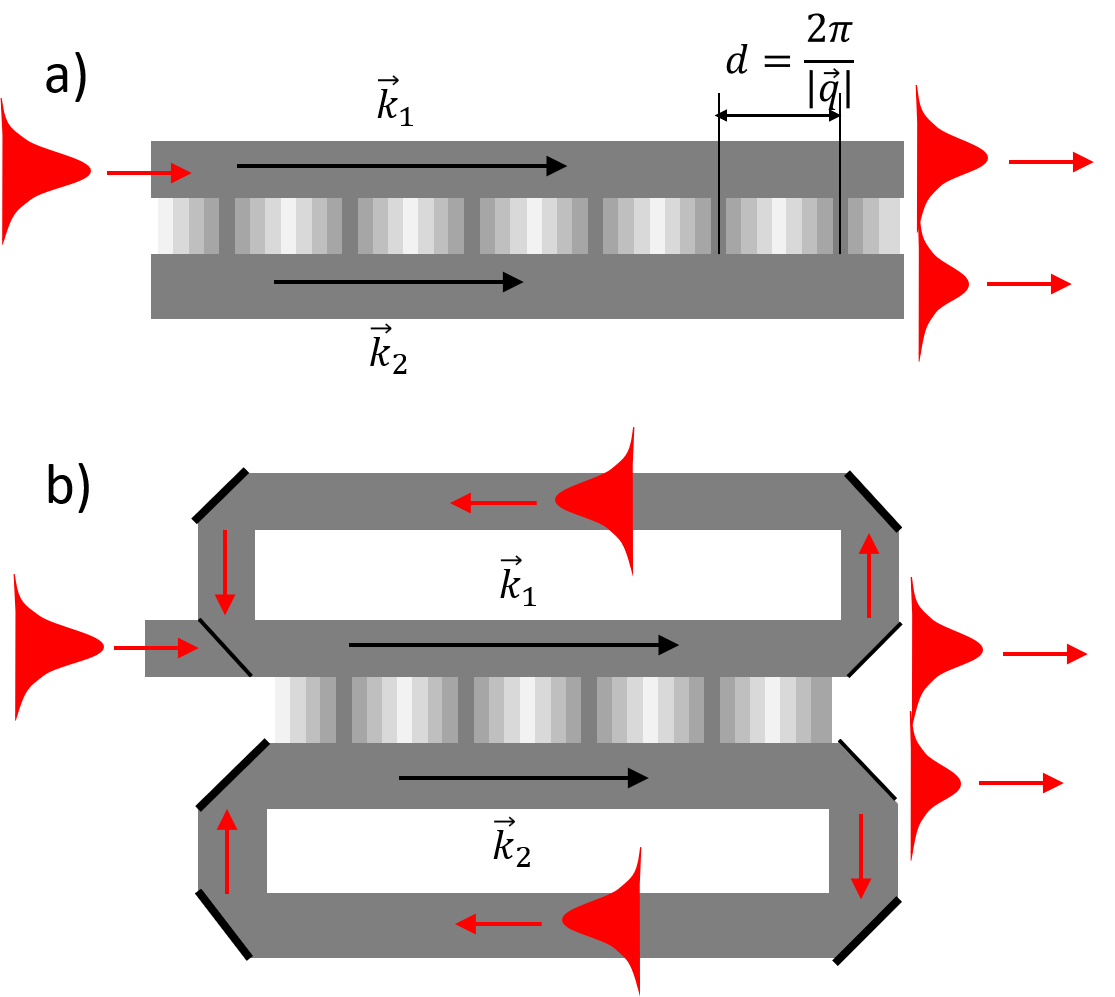}
    \caption{Quasi-PM coupling in autonomous a), and driven systems b). In autonomous case the radiation is injected into waveguide 1 through initial conditions, and is coupled to the waveguide 2. Evolution occurs along the waveguides (in z space). In driven case radiation is permanently injected into resonator 1 and is coupled to resonator 2. Evolution occurs in time t. In both cases coupling provides quasi-phase matching between different propagation wavevectors  $k_1$ and  $k_2$. }
    \label{fig:Picture1}
\end{figure}

Recently, unidirectional coupling has been proposed by unidirectional quasi-phase matching (quasi-PM)  \cite{Ivars2022,Akhter2024}. The simplest physical picture is the following. Consider two single-mode waveguides supporting guided waves with propagation constants $k_1$ and $k_2$, coupled by a periodic modulation of the interface with wave number $q\simeq k_1-k_2$, see Fig.~\ref{fig:Picture1}(a).  A conventional Hermitian coupling potential of the potential barrier (coupling interface), proportional to $\cos qx$, contains both Fourier components $e^{iqx}$ and $e^{-iqx}$, and therefore couples the two waves reciprocally. In contrast, if the coupling potential of the barrier were proportional, for instance, only to $e^{iqx}$, i.e., a ``half cosine,'' this would couple unidirectionally. In real space, such a complex modulation can be synthesized by imposing the real and imaginary quadratures with equal amplitudes and a relative phase shift of $\pi/2$:
\begin{equation}
\epsilon(x)=\epsilon_{\mathrm r}\cos(qx)+i\epsilon_{\mathrm i}\cos(qx+\phi),
\label{eq:epsdef}
\end{equation}
with $\epsilon_{\mathrm r}=\epsilon_{\mathrm i}$ and $\phi=\pi/2$ (or $3\pi/2$).

This line of reasoning naturally suggests that maximal unidirectionality should occur when two conditions are simultaneously satisfied: (i) the phase shift between the reactive and dissipative modulations is exactly $\pi/2$, i.e., the modulation is $\mathcal{PT}$-symmetric; and (ii) the amplitudes of the real and imaginary modulations are balanced, i.e., the system is precisely at an exceptional point (EP) of the effective two-mode model. The literature on $\mathcal{PT}$-symmetric Bragg structures, unidirectional invisibility, and exceptional-point photonics further reinforces this expectation \cite{Lin2011,Regensburger2012,Feng2013,Bender1998,Bender2007,ElGanainy2018,Ozdemir2019,Miri2019}.

However, recent studies on several, apparently different systems indicate that the actual situation is more subtle. In autonomous systems non-Hermitian quasi-PM modulations were shown to induce directional energy transfer between internal modes of multimode waveguides and thus provide non-Hermitian mode cleaning or turbulence control \cite{Ivars2022,Akhter2024}. This mathematically is equivalent to the case shown in Fig.~\ref{fig:Picture1}(a). (Fig.~\ref{fig:Picture1}(a) the case between two waveguides, whereas in \cite{Akhter2024} mathematically equivalent coupling between two different modes of the same waveguide was considered.) A different situation was realized in \cite{Grineviciute2025} where asymmetric in-coupling and out-coupling between incoming radiation and guided modes gives rise to non-Hermitian light trapping at Fano resonances. This is a non-autonomous case, and is illustrated in Fig.~\ref{fig:Picture1}(b). In both cases the maximal unidirectionality effect occurs at the phase shift between the real and imaginary quadratures of complex index modulation different from $\phi=\pi/2$ (or from $3\pi/2$).

The purpose of the present work is to clarify this contradiction. The main message is simple: In an autonomous system the relevant measure of asymmetry is determined by an interplay between the asymmetry of the eigenvector and the growth rate of the dominant eigenmode. At the EP the eigenvector is extremely asymmetric, but the growth exponent vanishes, and therefore the observable asymmetry is undefined. In a driven system the response is not selected dynamically by modal growth; it is obtained directly from the resolvent of the linear operator. Hence the observable is governed by poles and zero-pole combinations of the response function itself. 

The paper is organized as follows. Section~II formulates the common physical setting and introduces the primary control parameters. Section~III addresses the autonomous case: the eigenvalues, the exact solution of the initial-value problem, the asymptotic participation ratio, and its normal forms. Section~IV treats the driven case, where the response amplitude exhibits pole and mixed zero-pole singularities. Section~V discusses the common geometric structure of the parameter space and highlights the difference between the two normal forms.

\section{Unidirectional coupling from non-Hermitian modulations}

The physical origin of unidirectionality is best understood in the Fourier representation. Let a complex periodic modulation $\epsilon(x)$ couple two resonant waves $\exp(ik_1x)$ and $\exp(ik_2x)$. The elements of the coupling matrix are proportional to the Fourier components of the modulation in the mismatched wave numbers $\pm (k_1-k_2)$. If both coupling harmonics are present with equal strength, then the two reciprocal mode-coupling channels are equivalent. If one harmonic is suppressed, one obtains asymmetric coupling. A compact way to encode this asymmetry is through the complex coupling coefficients
\begin{equation}
\kappa_{1,2}=
\frac{1}{2}
\qty(
\epsilon_{\mathrm r}+
i\epsilon_{\mathrm i} e^{\pm i\phi})
\label{eq:betapm}
\end{equation}

The corresponding exceptional points $\mathrm{EP}_1$ and $\mathrm{EP}_2$, are defined by $\kappa_1=0$ and $\kappa_2=0$, respectively, which are realized in parameter space at $\epsilon_{\mathrm r}=\epsilon_{\mathrm i}$, and at $\phi=\pi/2$,  and at $\phi=3\pi/2$ respectively. These two points are the two simplest exceptional points of the reduced two-mode problem.

Equation~\eqref{eq:betapm} also makes the relation to spatial Kramers-Kronig media explicit. If the modulation is proportional to $e^{iqx}$, then its real and imaginary parts are Hilbert transforms of each other. In the present problem, however, the object of interest is not the scattering of a continuum from a generic Kramers-Kronig medium but the resonant quasi-phase-matched coupling between a small number of selected modes. This restriction is precisely what allows the full problem to be compressed to a small matrix model and analyzed in closed form. At the same time, because the interaction is resonant, the detuning from exact matching becomes an essential parameter in addition to the two modulation parameters $\phi$ and $\Delta\epsilon\equiv \epsilon_{\mathrm i}-\epsilon_{\mathrm r}$.

Therefore the natural control space is three-dimensional, $(\phi,\Delta\epsilon,\delta)$, where $\delta$ denotes the detuning from the exact resonance condition. In the autonomous case we write $\delta\equiv \Delta k$, while in the driven case we use the frequency detuning $\delta\equiv \Delta\omega$. The key result of the paper is that this is the same parameter space for both classes of systems, but the singular object that determines the optimum is different.

\section{Autonomous systems}

\subsection{Two-mode evolution problem}

For an autonomous system we consider two mode amplitudes $A_1(z)$ and $A_2(z)$, collected into the vector $\bm A=(A_1,A_2)^T$. Their evolution along the propagation coordinate $z$ obeys
\begin{equation}
\frac{d}{dz}
\begin{pmatrix}A_1\\A_2
\end{pmatrix}=i
\begin{pmatrix}
-\frac{\Delta k}{2} & \kappa_1 \\
\kappa_2 & \frac{\Delta k}{2}
\end{pmatrix}
\begin{pmatrix}A_1\\A_2
\end{pmatrix}.
\label{eq:auto1}
\end{equation}
Here $\Delta k$ is the mismatch from exact quasi-phase matching. The eigenvalues of the traceless evolution matrix are:

\begin{equation}
\lambda_{\pm}=\pm\lambda,
\qquad
\lambda=\sqrt{-\frac{\Delta k^2}{4}-\kappa_1\kappa_2},
\label{eq:lambda}
\end{equation}

and the branch with $\Re\,\lambda>0$ is used for the asymptotic analysis.

In the autonomous configuration considered below, radiation is injected into waveguide 1, and the quantity of interest is the normalized intensity transferred to waveguide 2. Therefore we impose  $
A_1(0)=1 $, and $
A_2(0)=0 $, and obtain the exact solution is
\begin{subequations}
\begin{align}
A_1(z)
&=
\cosh(\lambda z)
-
\frac{i\Delta k}{2\lambda}\sinh(\lambda z),
\frac{i\Delta k}{2\lambda}\sinh(\lambda z),
\label{eq:A1exact}
\\[1mm]
A_2(z)
&=
\frac{i\kappa_2}{\lambda}\sinh(\lambda z).
\label{eq:A2exact}
\end{align}
\end{subequations}
Consequently, the large-$z$ amplitude ratio for transfer into the second waveguide is
\begin{equation}
R_\infty
=
\lim_{z\to\infty}
\abs{\frac{A_2(z)}{A_1(z)}}
=
\abs{\frac{i\kappa_2}{\lambda-i\Delta k/2}}
=
\abs{\frac{\lambda+i\Delta k/2}{i\kappa_1}}.
\label{eq:auto2}
\end{equation}
The normalized second-mode intensity, which is the quantity plotted in the autonomous maps, is
The normalized second-mode intensity is
\begin{equation}
P_2(z)
=
\frac{|A_2(z)|^2}{|A_1(z)|^2+|A_2(z)|^2}.
\label{eq:P2}
\end{equation}
Equivalently, if $R(z)=|A_2(z)/A_1(z)|$, then $P_2(z)=R^2(z)/[1+R^2(z)]$, and asymptotically
\begin{equation}
P_{2,\infty}
=
\frac{R_\infty^2}{1+R_\infty^2}.
\label{eq:P2inf}
\end{equation}
This point is crucial because the spectral quantities alone are misleading. At either exceptional point one of the couplings becomes unidirectional and the eigenvector becomes formally maximally asymmetric. At the same time, however, the real part of the corresponding growth exponent can vanish. Thus the exceptional point is singular, but not necessarily optimal from the point of view of useful transfer into waveguide 2. Equation~\eqref{eq:auto2} already contains the required compromise between coupling asymmetry and modal growth.

At finite propagation distance one may still use the exact normalized intensity $P_2(z)$ to quantify mode cleaning. For the asymptotic maps and for the normal-form analysis, Eq.~\eqref{eq:auto2} is the more transparent quantity because it isolates the branch that survives at large $z$.

\subsection{Autonomous eigenmodes and mode-cleaning maps}

\begin{figure}[t]
    \centering
    \includegraphics[width=1\columnwidth]{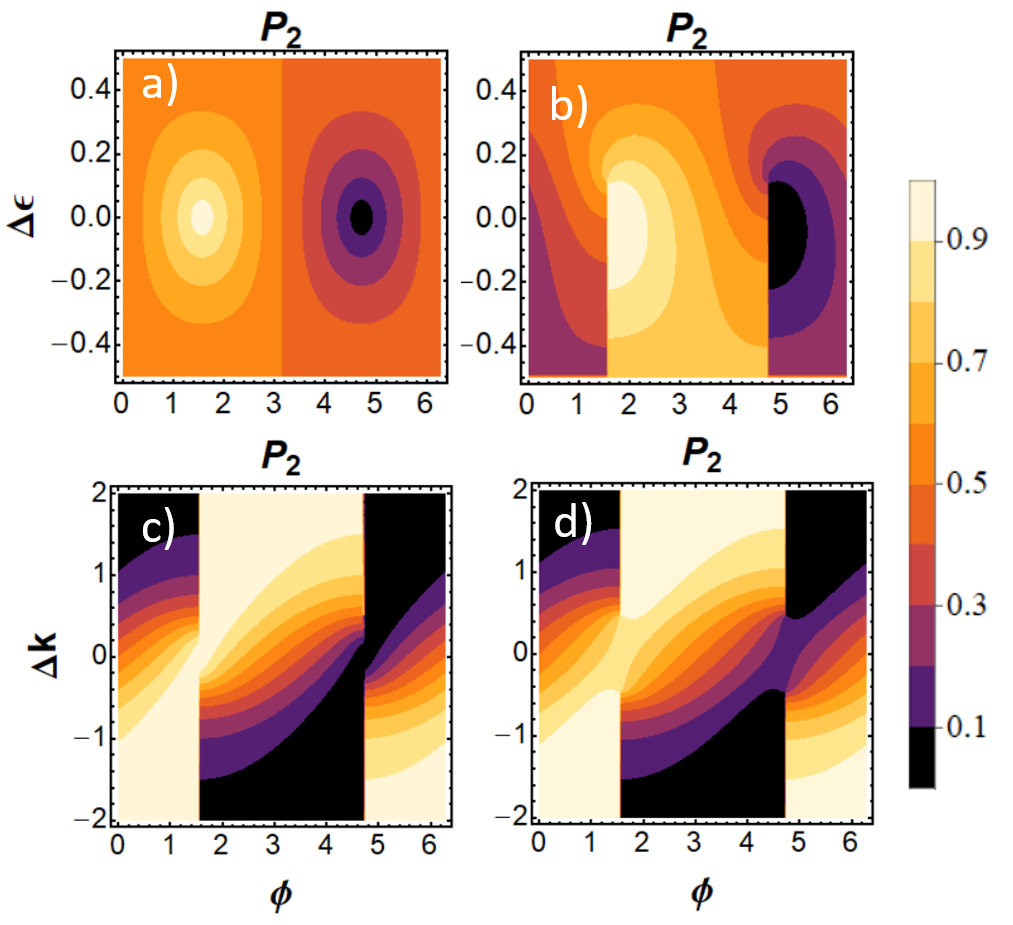}
    \caption{Autonomous regime: asymptotic mode-cleaning maps. Top row: second mode participation ratio depending on phase shift and amplitude imbalance $\Delta\epsilon$ at resonance  $\Delta k=0$ (a), and out of resonance  $\Delta k=0.5$ (b). Bottom row: second mode participation ratio depending  on phase mismatch $\Delta k$ at balanced index and gain/loss modulation $\Delta\epsilon=0$ (c), and imbalanced  $\Delta\epsilon=0.2$  (d). }
    \label{fig:Picture2}
\end{figure}

Before turning to the normal forms, it is helpful to summarize the behavior of the exact solution. Figure~\ref{fig:Picture2} collects representative maps of the normalized second-mode intensity $P_2$, calculated from the definition of $P_2$. This bounded observable avoids the infinities of the amplitude ratio and directly measures the fraction of power transferred from waveguide 1 to waveguide 2. 

The upper panels show $P_2$ in the $(\phi,\Delta\epsilon)$ plane at exact resonance and at finite mismatch, whereas the lower panels show $P_2$ in the $(\phi,\Delta k)$ plane for balanced and imbalanced real/imaginary modulation quadratures. The strongest transfer occurs in the vicinity of the two PT exceptional points $\mathrm{EP}_1$ and $\mathrm{EP}_2$, but the exceptional point itself is singular rather than optimal. Finite mismatch shifts the optimum systematically away from the nominal PT condition.

The physical reason for this shift is as follows: If one asks only which modulation profile makes one coupling coefficient vanish, the answer is the PT exceptional point. But if one asks which modulation profile makes the asymptotic population of one mode largest, the answer must also include the growth rate. Exactly at either EP the eigenvector is extremely asymmetric, yet the two competing branches become marginal. Hence the actual optimum is located where the coupling remains strongly asymmetric while the dominant mode still grows efficiently.

\begin{figure}[t]
    \includegraphics[width=1\columnwidth]{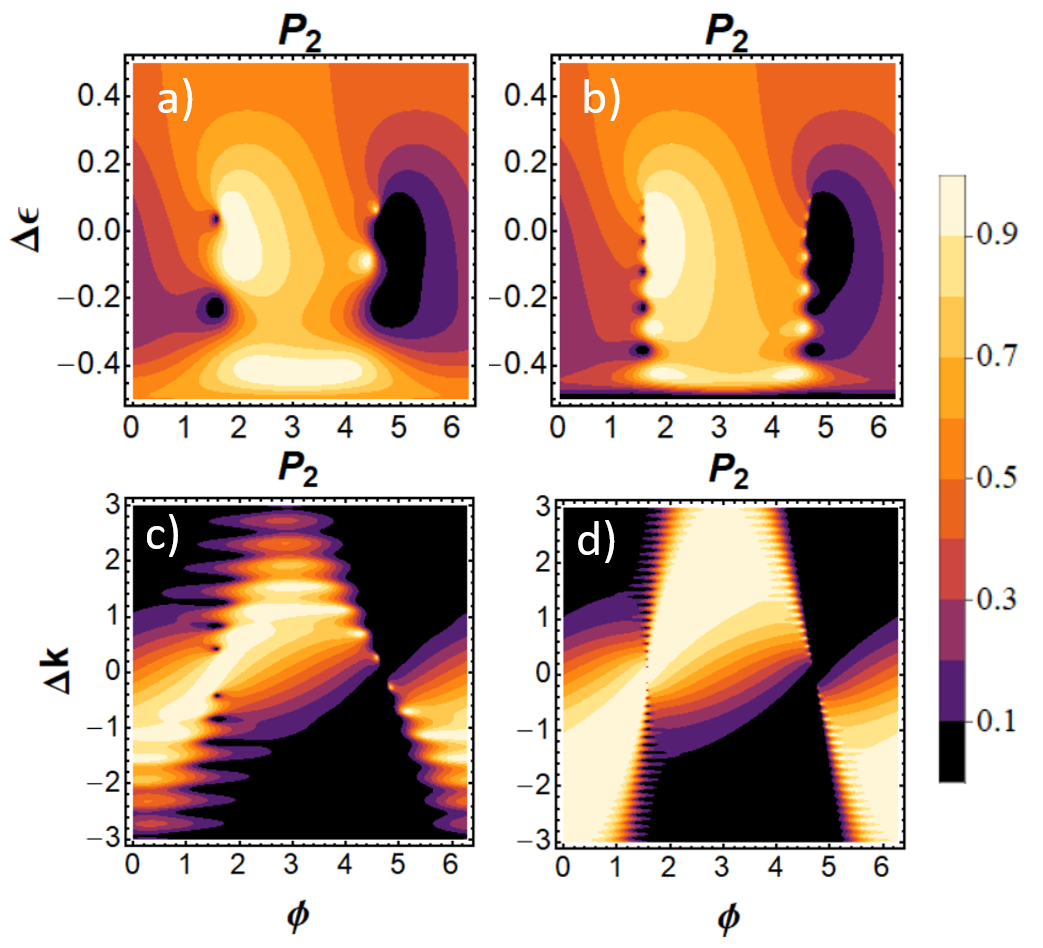}
    \caption{Autonomous regime at finite propagation distance $z$. Top row: second mode participation ratio depending on phase shift and modulation imbalance $\Delta\epsilon$ at propagation distances $z=15$ (a), and  $z=45$ (b). Off resonance, $\Delta k=0.5$. Bottom row: second mode participation ratio depending on phase shift and phase mismatch $\Delta k$ at propagation distances $z=15$ (c), and  $z=40$ (d). Modulation amplitudes are balanced $\Delta\epsilon=0$}
    \label{fig:Picture3}
\end{figure}

The finite interaction distance behaviour is shown in Fig.~\ref{fig:Picture3}. At moderate propagation distances the maps display alternating bright and dark regions because the evolution is oscillatory and the asymptotic selection of the dominant mode is incomplete. Increasing the propagation distance broadens the region where the second mode dominates. At first sight this broadening may suggest a nonresonant character of mode cleaning. However, the lower panels of Fig.~\ref{fig:Picture3}, which display the second-mode intensity itself, show that the largest absolute intensities remain concentrated near the resonance. Far from resonance a large participation ratio may still occur, but only because both amplitudes are small and the ratio ceases to be a good measure of useful conversion. Thus the practical optimum remains close to the resonant region.

\subsection{Normal forms near the exceptional points}

We now derive the local normal forms that govern the autonomous problem. Introduce the mean modulation strength $\epsilon_0\equiv (\epsilon_{\mathrm r}+\epsilon_{\mathrm i})/2$, the imbalance $\Delta\epsilon\equiv \epsilon_{\mathrm i}-\epsilon_{\mathrm r}$, and the local phase variables
\[
\Delta\phi_1\equiv \phi-\frac{\pi}{2},
\qquad
\Delta\phi_2\equiv \phi-\frac{3\pi}{2}.
\]
To leading order in $\Delta\epsilon$, $\Delta\phi_{1,2}$, and $\Delta k$, one finds from Eq.~\eqref{eq:betapm} that
\[
\begin{aligned}
\kappa_1\kappa_2\big|_{\phi\approx \pi/2}
&\simeq -\epsilon_0\left(\frac{\Delta\epsilon}{2}+\frac{i\epsilon_0}{2}\Delta\phi_1\right),
\\
\kappa_1\kappa_2\big|_{\phi\approx 3\pi/2}
&\simeq +\epsilon_0\left(\frac{\Delta\epsilon}{2}-\frac{i\epsilon_0}{2}\Delta\phi_2\right).
\end{aligned}
\]

Substituting these expressions into the definition of $\lambda$ yields the cleaned eigenvalue normal forms
\begin{equation}
\lambda_{1,2}
\simeq
\sqrt{
-\frac{\Delta k^2}{4}
\pm \frac{\epsilon_0\Delta\epsilon}{2}
+i\frac{\epsilon_0^2}{2}\Delta\phi_{1,2}
},
\label{eq:auto7}
\end{equation}
where the upper and lower signs correspond to the neighborhoods of $\mathrm{EP}_1$ and $\mathrm{EP}_2$, respectively.
These expressions display the expected square-root branching of an exceptional point. The branch point is shifted by the phase mismatch from the nominal PT point to a displaced position in the complex $(\phi,\Delta\epsilon)$ plane. This is the first manifestation of the shifted optimum.

To obtain the experimentally relevant normal form one must substitute Eq.~\eqref{eq:auto7} into the asymptotic transfer ratio~\eqref{eq:auto2}. Around $\mathrm{EP}_1$ this gives
\begin{equation}
R_{\infty,1}\simeq \abs{\frac{\epsilon_0}{\lambda_1-i\Delta k/2}},
\qquad
R_{\infty,2}\simeq \abs{\frac{\lambda_2+i\Delta k/2}{\epsilon_0}}.
\label{eq:auto8}
\end{equation}
with the second expression describing the neighborhood of $\mathrm{EP}_2$.
The point to stress is not the exact prefactor but the singular structure: near $\mathrm{EP}_1$ the transfer ratio has a pole-like form, while near $\mathrm{EP}_2$ it behaves as a zero attached to the square-root branch of the eigenvalue. This is why the maps of Fig.~\ref{fig:Picture2} are strongly asymmetric once detuning is introduced.

This distinction is physically important. The eigenvalues themselves always show the classic EP square-root structure. But the experimentally relevant modal ratio is not the eigenvalue; it is a nonlinear function of the eigenvalue and the couplings. Therefore the effective normal form of the observable is more involved than the square-root topology of the spectral problem alone. This is precisely why looking only at the EP of the coupling matrix can be misleading when trying to optimize mode cleaning.

\section{Driven systems}

\subsection{Two-mode response model}

We now consider an externally driven configuration. The prototype is a corrugated thin film supporting a guided mode that is fed from free space through a non-Hermitian modulation of the interface. In this situation the key issue is not the competition of internal growth rates but the strength of the forced response. To keep the notation parallel to the autonomous case, we denote by $A_1(t)$ the driven leaky or Fabry-P\'erot-like amplitude and by $A_2(t)$ the guided-mode amplitude. The equations are
\begin{equation}
\frac{d}{dt}
\begin{pmatrix}A_1\\A_2\end{pmatrix}=
\begin{pmatrix}
-\tau^{-1}-\gamma_0 & i\kappa_+\\
i\kappa_- & -\gamma-i\Delta\omega
\end{pmatrix}
\begin{pmatrix}A_1\\A_2\end{pmatrix}+
\begin{pmatrix}A_{\rm in}/\tau\\0\end{pmatrix}.
\label{eq:driven1}
\end{equation}
Here $\tau$ is the lifetime of the driven continuum-like mode, $\gamma_0$ and $\gamma$ denote background losses, $\Delta\omega$ is the detuning between the drive and the guided mode, $\kappa_\pm$ are the two non-Hermitian coupling coefficients defined in Eq.~\eqref{eq:betapm}, and $A_{\rm in}$ is the incident driving amplitude. In the stationary regime one solves a $2\times2$ linear system and obtains
\begin{equation}
A_2=
\frac{i\kappa_- A_{\rm in}/\tau}
{(\tau^{-1}+\gamma_0)(\gamma+i\Delta\omega)+\kappa_+\kappa_-}.
\label{eq:driven2}
\end{equation}
The crucial point is that the observable is directly a rational function of the control parameters.

In contrast with the autonomous case, there is no modal selection mechanism here. The amplitude $A_2$ is determined directly by the inverse of the linear operator. Therefore, near the exceptional points, one expects pole singularities rather than an interplay between branch topology and growth competition.

\subsection{Fano-type response maps and shifted poles}

\begin{figure}[t]
    \includegraphics[width=\columnwidth]{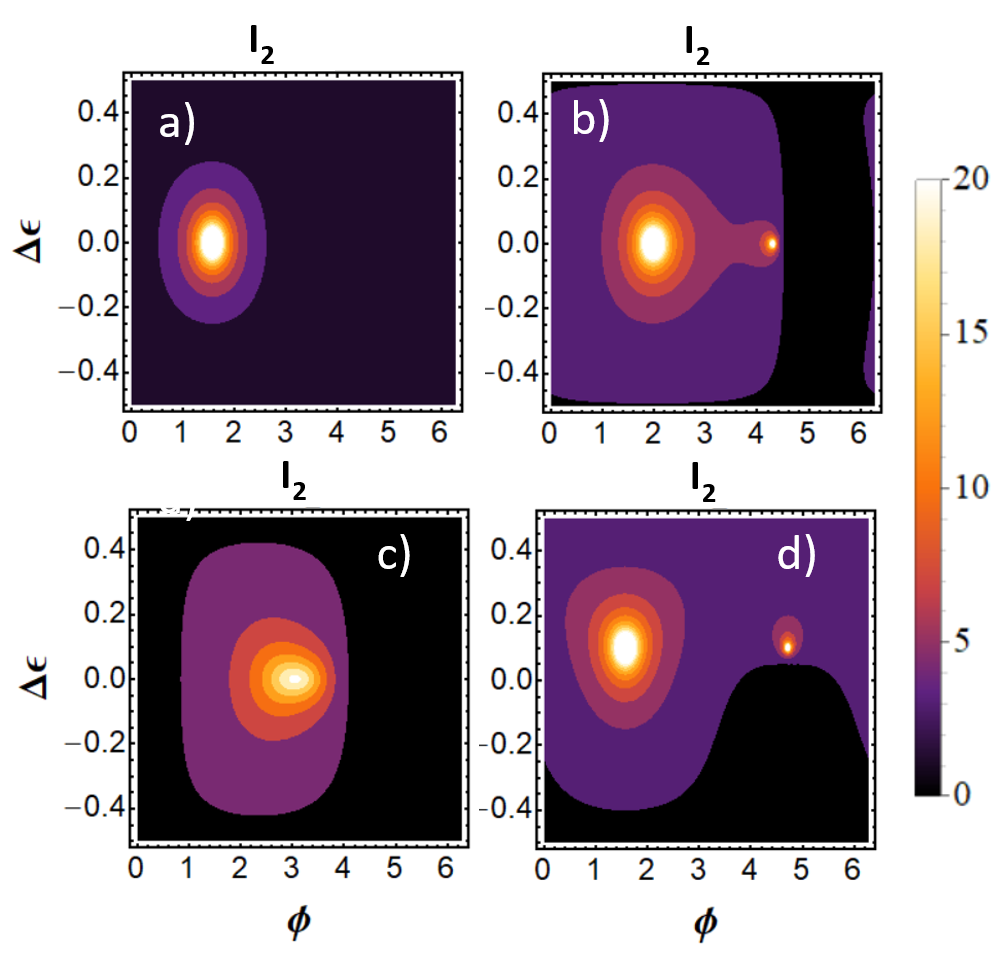}
    \caption{Driven case: response-amplitude maps in the $(\phi,\Delta\epsilon)$ plane. The principal singularity is located at the EP for zero detuning $\Delta\omega=0$ (a) and is shifted away from the EP for nonzero detuning $\Delta\omega=-0.05$ (b). In the latter case a secondary weaker zero-pole singularity appears near the second EP. With increasing detuning $\Delta\omega=-0.15$ the two singular structures approach one another and eventually merge (c). Losses shift the singularities in the vertical direction, shown for $\gamma=0.05$ and $\Delta\omega=0$ (d).}
    \label{fig:Picture4}
\end{figure}

\begin{figure}[t]
    \includegraphics[width=\columnwidth]{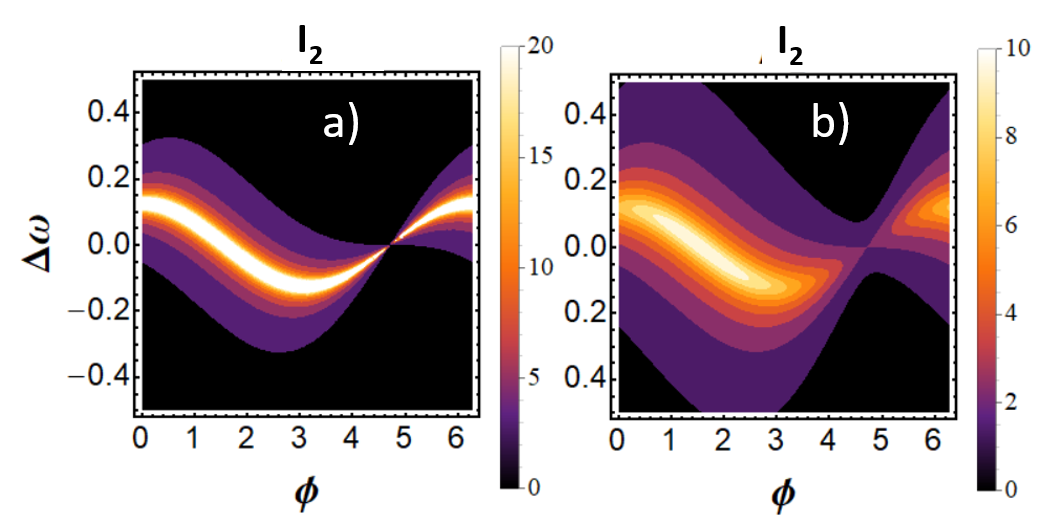}
    \caption{Driven case: response amplitude in the $(\phi,\Delta\omega)$ plane. Different phase shifts correspond to different detuning for maximal excitation, which makes the singular set a shifted ridge rather than a single point. This explains why the optimum unidirectionality is displaced from the PT-symmetric condition. In gain/index balanced case $\Delta\epsilon=0$ (a), and imbalanced case $\Delta\epsilon=0.2$ (b).}
    \label{fig:Picture5}
\end{figure}

Representative maps of the response amplitude are shown in Figs.~\ref{fig:Picture4} and \ref{fig:Picture5}. At exact resonance and in the absence of additional losses, the response forms a pole centered near one of the exceptional points, as anticipated from the basic PT intuition. However, a finite detuning $\Delta\omega$ shifts the pole away from the nominal EP. Simultaneously, a second weaker singularity appears near $\mathrm{EP}_2$. For increasing detuning the two singular structures approach each other and eventually merge in the anti-PT sector, i.e., near in-phase or antiphase modulations rather than at a quadrature phase shift.

This behavior, which at first seems surprising, is in fact completely natural once one remembers that the stationary response is governed by the denominator of Eq.~\eqref{eq:driven2}. The EP is the point where one coupling vanishes, but the maximum response occurs where the full denominator is minimized. The detuning enters this denominator on the same footing as the modulation parameters and displaces the singular point. Background losses add yet another shift, this time in a direction approximately orthogonal in the $(\phi,\Delta\epsilon)$ plane. Thus the driven problem does not probe the exceptional point itself; it probes the pole of the resolvent, which is displaced from the EP.

The same structure becomes even clearer when the response is plotted in the $(\phi,\Delta\omega)$ plane at fixed modulation amplitudes, as in Fig.~\ref{fig:Picture5}. At balanced modulation amplitudes, one obtains a continuous singular ridge: different phase shifts require different detunings for maximal excitation. This singular ridge is the driven-system counterpart of the shifted optimum in the autonomous problem. When the amplitudes become unbalanced, the two singular branches no longer have equal weight and the response becomes strongly skewed. In particular, the singularity associated with the second exceptional point is much weaker for small detuning and grows only as the two branches approach coalescence.

\subsection{Normal forms in the driven problem}

The local expansions around the two exceptional points are particularly transparent in the driven system. In terms of the compact notation $\kappa_\pm$, the two branches are still generated by the vanishing of one coupling at a time; we keep the labels $1,2$ in the formulas below because they map directly onto the two physical poles seen in the figures. Using the same variables $\epsilon_0$, $\Delta\epsilon$, $\Delta\phi_1$, and $\Delta\phi_2$ as in the autonomous case, one finds from Eq.~\eqref{eq:betapm} that
\begin{subequations}
\begin{align}
\kappa_+\big|_{\phi\approx \pi/2}&\simeq -\frac{\Delta\epsilon}{2}-i\frac{\epsilon_0}{2}\Delta\phi_1,
\qquad
\kappa_-\big|_{\phi\approx \pi/2}\simeq \epsilon_0,
\label{eq:betadr1}\\
\kappa_-\big|_{\phi\approx 3\pi/2}&\simeq -\frac{\Delta\epsilon}{2}+i\frac{\epsilon_0}{2}\Delta\phi_2,
\qquad
\kappa_+\big|_{\phi\approx 3\pi/2}\simeq \epsilon_0.
\label{eq:betadr2}
\end{align}
\end{subequations}
Substituting these expansions into Eq.~\eqref{eq:driven2} and retaining the leading terms gives
\begin{subequations}\label{eq:driven3}
\begin{align}
A_2\big|_{\phi\approx \pi/2}
&\simeq \frac{i\epsilon_0 A_{\rm in}/\tau}{(\tau^{-1}+\gamma_0)(\gamma+i\Delta\omega)-\epsilon_0\Delta\epsilon/2-i\epsilon_0^2\Delta\phi_1/2},
\label{eq:driven3a}\\
A_2\big|_{\phi\approx 3\pi/2}
&\simeq \frac{iA_{\rm in}\left(-\Delta\epsilon/2+i\epsilon_0\Delta\phi_2/2\right)/\tau}{(\tau^{-1}+\gamma_0)(\gamma+i\Delta\omega)-\epsilon_0\Delta\epsilon/2+i\epsilon_0^2\Delta\phi_2/2}.
\label{eq:driven3b}
\end{align}
\end{subequations}
These are the central normal forms for the driven problem. Equation~\eqref{eq:driven3a} is a genuine complex pole in the two-dimensional control space $(\Delta\phi_1,\Delta\epsilon)$. Equation~\eqref{eq:driven3b}, in contrast, contains a zero in the numerator and a pole in the denominator. This is a zero-pole composite singularity. Unlike Eq.~\eqref{eq:driven3a}, it is not a genuine pole because the same small parameter that drives the denominator toward zero simultaneously suppresses the numerator. Such mixed zero-pole singularities appear to be uncommon in non-Hermitian optics and may deserve further mathematical study.

If background losses are neglected for simplicity, the displacement of the singularities can be obtained directly from the zero of the denominator. For the first branch, Eq.~\eqref{eq:driven3a}, one finds
\begin{equation}
\frac{\epsilon_0\Delta\epsilon}{2}
+i\frac{\epsilon_0^2\Delta\phi_1}{2}
\simeq
i\frac{\Delta\omega}{\tau}.
\label{eq:shiftclean}
\end{equation}
Thus the singular point is shifted away from the nominal EP by
\begin{equation}
\Delta\phi_1
=
-\frac{2\Delta\omega}
{\tau\epsilon_0^2}.
\label{eq:shift1}
\end{equation}
Similarly, for the second branch one obtains
\[
\Delta\phi_2
=
+\frac{2\Delta\omega}
{\tau\epsilon_0^2},
\qquad
\Delta\phi_{\rm sep}
=
\frac{4\Delta\omega}
{\tau\epsilon_0^2}.
\]
The separation between the two shifted singularities is therefore given by the second expression above.
As the detuning increases, the two singular branches move toward each other. They coalesce when the detuning-induced displacement becomes comparable to the angular separation of the two nominal exceptional points,
\begin{equation}
|\Delta\omega|
=
\frac{\pi\epsilon_0^2\tau}{4}.
\label{eq:coalescence}
\end{equation}
This prediction is directly confirmed by the numerical maps: the two shifted maxima approach each other approximately linearly with increasing detuning and merge at the value given by Eq.~\eqref{eq:coalescence}. The observed coalescence provides an independent numerical validation of the normal-form description. Background losses add a constant complex offset to the right-hand side of Eq.~\eqref{eq:shiftclean}, displacing the singularities further in the $(\phi,\Delta\epsilon)$ plane. This is the driven analogue of the shifted optimum found in the autonomous problem.

\section{Conclusions}

We have studied unidirectional coupling in quasi-phase-matched non-Hermitian systems in both autonomous and driven regimes. Although physically different, both systems originate from the same mechanism: suppression of one resonant Fourier component by a complex modulation profile, leading to asymmetric coupling and exceptional-point physics.

A central result is that the experimentally relevant extrema generally do not coincide with the nominal exceptional points. Detuning, and in driven systems also background gain or loss, shift the singular structures away from the ideal $\mathcal{PT}$-symmetric condition. The resulting displacements can be obtained analytically and accurately reproduce the numerical maps.

The autonomous and driven systems possess different normal forms, yet display remarkably similar singular behavior. In both cases exceptional points organize the geometry of parameter space, while the optimum response is shifted away from them. The driven problem additionally reveals a mixed zero-pole singularity, where numerator and denominator simultaneously approach zero. Exceptional points should therefore be viewed primarily as organizing singularities rather than universal operating points of maximum performance.

\begin{acknowledgments}
This project has received funding from Spanish Ministry of Science, Innovation and Universities (MICINN) under the project PID2022-138321NB-C21.
\end{acknowledgments}

\end{document}